\documentclass[preprint,preprintnumbers,amsmath,amssymb,nofootinbib]{revtex4}
\usepackage{amsmath,amssymb,hyperref}
\usepackage{graphicx}
\usepackage{dcolumn}
\usepackage{bm}

\newcommand{\Alpha}{\mathrm{A}}
\newcommand{\mathd}{\mathrm{d}}
\newcommand{\tmem}[1]{{\em #1\/}}
\newcommand{\tmop}[1]{\ensuremath{\operatorname{#1}}}

\newcommand{\tmtextbf}[1]{{\bfseries{#1}}}
\newcommand{\tmtextit}[1]{{\itshape{#1}}}
\newcommand{\tmtexttt}[1]{{\ttfamily{#1}}}

\begin{document}

\title{Hawking Radiation in the Ghost Condensate \\ is Non-Thermal}

\author{Brian Feldstein}
 \email{bfeldste@bu.edu}
\affiliation{%
Department of Physics\\
Boston University\\
Boston, MA, 02215}%

\begin{abstract}
We consider a Schwarzschild black hole immersed in a ghost
condensate background. \ It is shown that the Hawking radiation in the quanta
of small perturbations around this background is highly suppressed- in
particular it is not given by a thermal spectrum. \ This result is in accord
with observations that such black holes can be used to violate the generalized
second law of thermodynamics, and thus \ cannot have a standard entropy/area
relation.
\end{abstract}

\maketitle

\section{Introduction}

The classic results of Bekenstein {\cite{Bekenstein}} and Hawking
{\cite{Hawking}} that black holes have an entropy proportional to their area
and that they radiate as blackbodies, have proven to be of fundamental
importance to our current understanding of quantum gravity. \ It is now widely
believed that any complete theory of quantum gravity must satisfy the
holographic principle {\cite{Hooft,Susskind}}, which states that the number of
fundamental degrees of freedom contained within a given region scales as the
area bounding that region, not as its volume. \ The holographic conjecture
itself was motivated by the noted discoveries in black hole thermodynamics,
and has gained further support from the AdS/CFT correspondence
{\cite{Maldacena}}.

It is very interesting in this context to try to find theories for
which black hole thermodynamics does {\tmem{not}} behave in the way expected
from holography. \ If we can find low energy effective theories with this
property, then we may be able to reach the striking conclusion that these
theories are incompatible with fundamental principles of quantum gravity.

An important example can be found in {\cite{Sergeis}}, in which
the authors studied a Schwarzschild black hole immersed in a ghost condensate
background {\cite{ghost}}. \ In the ghost condensate, a scalar field has a
non-vanishing expectation value for its kinetic term, while having essentially
zero energy-momentum. \ This background breaks Lorentz invariance, and adding
couplings between the ghost condensate and additional fields, we can obtain
particles with maximum speeds which differ from the speed of light.

In the black hole background, such particles have horizons at
different radii than the usual horizon, and by extension, Hawking spectra with
temperatures different than the usual temperature. \ Of course, already the
thermodynamic picture of such a black hole is a bit mysterious- it is not
clear for example what temperature (or entropy) is the ``correct'' one. \
Indeed, the authors of {\cite{Sergeis}} showed that it is possible to use such
a black hole and an arrangement of shells at various temperatures to violate
the generalized second law of thermodynamics ({\cite{Jacobson3}} used an
alternative mechanism to reach a similar conclusion)\footnote{This author is
somewhat concerned that the perpetual motion machine discussed in
{\cite{Sergeis}} (and also the one in {\cite{Jacobson3}}) has a possible
loophole, in that the flux of particles heading into the black hole may cause
a relevant shift in the ghost condensate background. This could lead to an
extra flux of ghost condensate energy into the black hole, and therefore also
to growth in the size of the horizon. This effect is at the right order of
magnitude to possibly (but not necessarily) ruin the claimed GSL violation. 
To analyze this issue at a greater level of precision will require further
study.}.

These facts could lead one to speculate that the ghost condensate
could not emerge in a low energy effective field theory coming from a
consistent quantum theory of gravity. \ In fact, a UV completion of any kind
for the ghost condensate has yet to be found.

It is then important to understand what exactly the fundamental
property of the ghost condensate theory is which causes its apparent
difficulties with quantum gravity. \ The likely candidate for such a property
is the violation of the null energy condition present in these
theories.{\footnote{The null energy condition is the requirement that, for any
future pointing null vector $e^{\mu}$, $T_{\mu \nu} e^{\mu} e^{\nu} \geqslant
0$, where $T \mu \nu$ is the energy-momentum tensor. \ Physically this can be
thought of as the requirement that light rays are always caused to be focused,
rather than anti-focused, by sources of energy.}} \ It is this feature which
allows the ghost condensate to break Lorentz invariance while simultaneously
carrying negligible energy-momentum- the key feature which allows the
construction of the perpetual motion machines of {\cite{Sergeis}} and
{\cite{Jacobson3}}.

It is interesting that, independent of any considerations
involving the null energy condition, the arguments of {\cite{Sergeis}} seem to
demonstrate a pathology with black hole thermodynamics in the ghost condensate
using only the properties of the Hawking radiation in various species. \ It
seems extremely interesting that one could compute a Hawking spectrum and
conclude that a particular low energy effective field theory is incompatible
with principles of quantum gravity.

With this motivation, in this paper we will attempt to complete
the story of the Hawking radiation in the ghost condensate by computing the
spectrum of the ghost condensate quanta themselves. \ Indeed, this will allow
us to see a pathology in the Hawking spectrum without having to require
specific sorts of couplings to additional fields: \ We will find that the
Hawking radiation in these particles is highly suppressed, and in particular,
is non-thermal. \ The radiation vanishes in the limit that $\frac{1}{2 M} \ll
\Lambda \ll M_{\tmop{pl}}$, where $M$ is the black hole mass, and $\Lambda$ is
the UV cutoff of the ghost condensate effective theory. \ Moreover, the flux
in a frequency $\omega$ vanishes in this limit even if we hold fixed the
product $\omega M$, thus ruling out a typical Boltzmann distribution such as
$e^{- \omega / T}$, with $T \propto 1 / M$. \ Note that here and throughout
the paper, we will work in natural units with $M_{\tmop{pl}} = 1$.

In section (\ref{2D}), we will review a useful calculation of \
the Hawking spectrum for a massless scalar in 2 dimensions, while in section
(\ref{GhostSection}) we will review relevant aspects of ghost condensation. \
In section (\ref{higher}) we will discuss the possible choices of higher
derivative operators in the theory, and their possible impact on our results.
\ In section (\ref{spectrum}) we will calculate the Hawking spectrum of ghost
condensate quanta, showing that it is non-thermal. \ We will conclude in
section (\ref{discussion}).

\section{\label{2D}Massless Scalar Hawking Spectrum in 2D}

Here we review a particular method for calculating the Hawking spectrum for a
massless scalar field in 1+1 dimensions {\cite{Jacobson1}}. \ This method will
later be shown to be generalizable to a calculation of the spectrum of ghost
condensate quanta. Much of our analysis follows the line of
argument in the excellent review \ by Jacobson {\cite{Jacobson2}}.

We consider a massless scalar field $\psi$ propagating in a
two-dimensional Schwarzschild black hole background. \ The metric is
\begin{equation}
  \tmop{ds}^2 = \left( 1 - \frac{2 M}{r} \right) \tmop{dt}^2 - \frac{1}{\left(
  1 - \frac{2 M}{r} \right)} d \underline{} r^2 . \label{2dmetric}
\end{equation}
We make a change of time coordinate to
\begin{equation}
  \tau \equiv t + 2 M \left( 2 \sqrt{\frac{r}{2 M}} + \log \left(
  \frac{\sqrt{r} - \sqrt{2 M}}{\sqrt{r} + \sqrt{2 M}} \right) \right),
  \label{tau}
\end{equation}
so that the metric takes the form
\begin{equation}
  \tmop{ds}^2 = \left( 1 - v^2 \right) \mathd \tau^2 + 2 v \tmop{drd} \tau -
  \tmop{dr}^2, \label{2dmetric2}
\end{equation}
where $v (r) \equiv - \sqrt{\frac{2 M}{r}}$. \ The velocity vector of an
observer freely falling into the black hole from rest at infinity is always
perpendicular to the surfaces of constant $\tau$. \ In fact $\tau$ measures
the proper time for such observers, and the $(\tau, r)$ coordinate system is
perfectly well behaved as one crosses the (future) event horizon. \ This
coordinate system is thus well suited to a calculation of the Hawking spectrum
generated at the horizon, and in fact it is particularly convenient for two
additional reasons: \ First, the time $\tau$ does not appear explicitly in the
metric, and second, as will be important later, the ghost condensate
background in the presence of a black hole satisfies $\varphi \propto \tau$.

 The action for the massless scalar is
\begin{equation}
  S = \int \sqrt{- g} g^{\mu \nu} \nabla_{\mu} \psi \nabla_{\nu} \psi .
  \label{2daction}
\end{equation}
In the metric (3) this takes the form
\begin{equation}
  S = \int \tmop{dr} \mathd \tau \left[ ((\partial_{\tau} + v \partial_r)
  \psi)^2 - (\partial_r \psi)^2 \right], \label{2daction2}
\end{equation}
and the equation of motion
\begin{equation}
  g^{\mu \nu} \nabla_{\mu} \nabla_{\nu} \psi = 0 \label{2deqnmot}
\end{equation}
takes the form
\begin{equation}
  \left[ \partial_{\tau}^2 + 2 v \partial_{\tau} \partial_r - (1 - v^2)
  \partial_r^2 + \frac{\partial v}{\partial r} \partial_{\tau} + 2 v
  \frac{\partial v}{\partial r} \partial_r  \right] \psi = 0.
  \label{2deqnmot2}
\end{equation}
As a result of the metric being independent of $\tau$, there exist solutions
to the equation of motion of the form $e^{- i \omega \tau} f (r)${\footnote{By
equation (2), these are also solutions of fixed killing frequency.}}. \ At
large distances from the hole, we may take $f (r)$ to approach$\sim e^{\pm
\tmop{ikr}}$, corresponding to outgoing or ingoing modes. \ A particle seen by
an observer at infinity may be taken to be a wave packet formed out of such
solutions peaked around some particular frequency of interest. Note that
purely outgoing modes will also have a boundary condition that they vanish
behind the horizon; without this boundary condition, the solution would also
describe a separate bit of matter falling into the singularity. We may then
ask, for example, about the expectation value of the number operator
corresponding to outgoing modes of this type, with a non-zero result
indicating particle production. \

 Now, associated with the wave equation (\ref{2deqnmot}), we have
an inner product defined on complex solutions given by
\begin{equation}
  < s_1, s_2 > \equiv i \int_{\Sigma} n^{\mu} [s_1^{\ast} \nabla_{\mu} s_2 -
  s_2 \nabla_{\mu} s_1^{\ast}] . \label{2dIP}
\end{equation}
Here $\Sigma$ is a given space-like hyper-surface, with $n^{\mu}$ being its
unit normal vector. \ Note that this inner product is not positive definite,
although it is independent of the surface chosen due to the Klein-Gordon
equation (\ref{2deqnmot}). \ On a surface of constant $\tau$, the inner
product takes the form
\begin{equation}
  < s_1, s_2 > = i \int \tmop{dr} [s_1^{\ast} (\partial_{\tau} + v \partial_r)
  s_2 - s_2 (\partial_{\tau} + v \partial_r) s_1^{\ast}] . \label{2dIP2}
\end{equation}
 Now, given a solution to the equation of motion $s$ we can define
an associated operator by
\begin{equation}
  a (s) \equiv < s, \Psi >, \label{a}
\end{equation}
with a Hermitian conjugate
\begin{equation}
  a^{\dag} (s) = - < s^{\ast}, \Psi > . \label{adagger}
\end{equation}
Here $\Psi$ is the quantum operator corresponding to $\psi$. \ It then follows
from the canonical commutation relations that
\begin{equation}
  [a (s_1), a^{\dag} (s_2)] = < s_1, s_2 > . \label{aalgebra}
\end{equation}
Thus if $s$ is a solution with positive norm in the inner product (\ref{2dIP})
(and we normalize it appropriately), then the operator $a (s)$ is in fact an
annihilation operator associated with the given mode. \ If $s$ has negative
norm, then its complex conjugate has positive norm and will be associated with
an annihilation operator.

 To compute the expectation value of the number operator associated
with a given solution $s$, we must thus evaluate $< a^{\dag} (s) a (s) >$. \
Note however that since the inner product (\ref{2dIP}) is conserved, we are
free to perform the computation on any space-like slice of our choosing. \
Exploiting this freedom, we may thus propagate an outgoing wave packet of
interest backwards in time until it is highly blue shifted and squeezed up
against the horizon. \ The reason this is fruitful is that, while we would
claim to have no a priori knowledge of the particle content of low energy
modes (such as the one we started with), we {\tmem{can}} say something about
the content of high energy modes: \ Since the evolution of the black hole
background is taking place on timescales of order $M$, modes with energy much
greater than $1 / M$ are expected to be in their ground states {\cite{nice}}.
\ Note that this argument applies to the energy as seen by observers who free
fall into the black hole from infinity. \ Equivalently, it applies so long as
the spacial slices we are using to define the time evolution and energy have
an extrinsic curvature which is always $\lesssim 1 / M$.

 We may thus propagate a low energy wave-packet of interest back in
time until it is squeezed against the horizon and composed primarily of high
energy modes as seen by a free falling observer. \ Note that such high energy
modes effectively propagate in flat space since their wavelengths are much
less than the Schwarzschild radius. \ In flat space, we have plane wave modes
of the form $e^{- i \omega t} e^{\pm i \omega x}$, and in the inner product
(\ref{2dIP}) these have positive norm for $\omega > 0$ and negative norm for
$\omega < 0$. \ When we evaluate $< a^{\dag} (s) a (s) >$on an early time
slice, we will thus decompose $s$ into {\tmem{positive and negative free fall
frequency}} pieces $s^{}_+$ and $s^{}_-$, with positive and negative norm. \
We will then have
\begin{equation}
  a (s) = < s, \Psi > = < s^{}_+, \Psi > + < s^{}_-, \Psi > = a (s^{}_+) -
  a^{\dag} (s_-^{\ast}), \label{olda}
\end{equation}
where in the last equality the positive and negative norms of $s_+$ and $s_-$
imply that $a (s^{}_+)$and \ $a^{\dag} (s_-^{\ast})$ are bona fide
annihilation and creation operators, respectively.{\footnote{More generally,
there is also a component at early times which is a positive norm, low energy,
ingoing mode, having been reflected off the geometry back out to infinity. \
For the wave-packets considered in this paper the reflected pieces will be
negligible.}}

 This allows us to calculate the expectation value of the number
operator we are interested in;
\begin{eqnarray}
  \text{$< a^{\dag} (s) a (s) >$} & = & < a (s_-^{\ast}) a^{\dag} (s_-^{\ast})
  > \nonumber\\
  & = & < s_-^{\ast}, s_-^{\ast} > \nonumber\\
  & = & - < s_-, s_- >,  \label{number}
\end{eqnarray}
where in the first line we have used the fact that the annihilation operators
for the high energy modes annihilate the vacuum, and in the second we have
used the commutation relation (\ref{aalgebra}). \ Thus the number operator for
a low energy outgoing wave-packet is given by the norm of its negative free
fall frequency component at early times. \ To find this negative free fall
frequency part, one may explicitly take components on a basis of modes using
the inner product (\ref{2dIP}). \ It is also possible to use a rather nice
trick due to Unruh {\cite{Unruh}} as we will shortly review.

 As mentioned above, we would like to form our wave-packet out of
solutions to the equation of motion of the form $e^{- i \omega \tau} f (r)$,
and which become outgoing plane waves at large distances. \ Now, it is
important to realize that although the coordinate $\tau$ does indeed measure
the proper time for free falling observers, modes of this form do {\tmem{not}}
have definite free fall frequency. \ This is a result of the existence of the
off diagonal term in the metric (\ref{2dmetric2}); derivatives in the
direction of a free falling observer's world-line are actually given by
$\partial_{\tau} + v \partial_r$ (by contrast, $\partial_r$ does take a
derivative in a direction perpendicular to this). \ Of course, a function of
the form $e^{- i \omega \tau} f (r)$ is not generally an eigenfunction of this
derivative operator.

 In flat space, our massless scalar satisfies the dispersion
relation $\omega^2 = k^2 .$ In the curved space-time of the black hole, we \
might hope to find solutions to the equation of motion by using a WKB
approximation. \ Such solutions take the form
\begin{equation}
  \text{$s_{\omega} (r) \propto e^{- i \omega \tau} e^{i \int^r k (r')
  \tmop{dr}'}$}, \label{WKB}
\end{equation}
with $k (r)$ being given by the {\tmem{local}} solution to the dispersion
relation at the given radius. \ That is, since \ \ \ $(\partial_{\tau} + v
\partial_r) s_{\omega} = - i (\omega - v k) s_{\omega}$, the ``local'' free fall
frequency is given by
\begin{equation}
  \omega_{\tmop{ff}} (r) = \omega - v (r) k (r), \label{wff}
\end{equation}
and we take
\begin{equation}
  \omega_{\tmop{ff}} (r)^2 = k (r)^2 . \label{2ddispersion}
\end{equation}
These WKB solutions will be good approximate solutions to the equation of
motion (\ref{2deqnmot2}) so long as the distance scales over which the
geometry and the wavelength change are much longer than the wavelength itself.
\ This corresponds to the requirements
\begin{equation}
  v' (r) \ll | k (r) v (r) |, \label{WKB1}
\end{equation}
and
\begin{equation}
  |k' (r) | \ll k (r)^2 . \label{WKB2}
\end{equation}
 In the case of the massless scalar, (\ref{wff}) and
(\ref{2ddispersion}) yield
\begin{equation}
  k (r) = \frac{\omega}{1 + v (r)}, \label{2dk}
\end{equation}
where we have chosen the solution which becomes an outgoing plane wave at
infinity. \ In fact for the massless scalar field, equations (\ref{WKB1}) and
(\ref{WKB2}) are {\tmem{not}} satisfied unless $\omega \gg 1 / M$, and one
would expect the WKB solutions to be invalid at wavelengths of interest of
order the Schwarzschild radius and larger. \ It turns out, however, that
solutions of the form (\ref{WKB}), with $k (r) \tmop{given} \tmop{by} (
\ref{2dk})$ just so happen to be {\tmem{exact}} solutions in the case of the
massless scalar in 1+1 dimensions (the terms in the equation of motion one
would want to drop by the WKB approximation actually happen to cancel each
other). \ The WKB method will still be useful in the case of the ghost
condensate, since in that case the analogues of equations (\ref{WKB1}) and
(\ref{WKB2}) {\tmem{will}} be valid.

 We now form a wave-packet $s$ out of the WKB modes, and propagate
it back in time until it is squeezed up against the horizon. \ We then wish to
find the positive and negative free fall frequency pieces. \ To this end we
thus only need to know the behavior of our modes close to the horizon, and we
expand $r = 2 M + x,$ finding at leading order
\begin{equation}
  k \simeq \frac{4 M \omega}{x}, \label{khorizon}
\end{equation}
and
\begin{equation}
  s_{\omega} \simeq e^{- i \omega \tau} e^{i 4 M \omega \tmop{Log} [x]} .
  \label{shorizon}
\end{equation}
The constant overall amplitude is irrelevant and has been dropped. \ These
expressions are valid for $x > 0$; recall that behind the horizon the outgoing
modes we are interested in vanish by definition. Now, an observer falling
freely into the black hole from infinity follows a path with $r = (
\frac{3}{2} \sqrt{2 M} (R - \tau))^{2 / 3}$ for some constant $R$. Close to
the horizon these paths have $x \simeq - \Delta \tau$, and thus the leading
behavior for the mode (\ref{shorizon}) along the free falling observer's
world-line is given by $s_{\omega} \simeq e^{i 4 M \omega \tmop{Log} [- \Delta
\tau]}$ for $\Delta \tau$ negative, with $s_{\omega} = 0$ for $\Delta \tau$
positive.

 Now the key thing to notice is that a mode formed as a linear
combination of phases $e^{- i \omega' \Delta \tau}$ with all $\omega'$
positive will be analytic and bounded in the {\tmem{lower}} half complex
$\Delta \tau$ plane. \ Conversely, if we take all the $w'$ negative the linear
combination will be analytic and bounded in the {\tmem{upper}} half complex
$\Delta \tau$ plane. \ Of course, neither such behavior is found for our mode
$s_{\omega}$ (which again, vanishes for positive $\Delta \tau$), indicating
that we have both positive and negative frequency components in our mode. \ On
the other hand, if we analytically extend the solution (\ref{shorizon}) to
negative $x$ values by putting a branch cut in either the upper or lower half
complex $x$ plane, then we {\tmem{will}} have purely positive or negative free
fall frequency solutions. \ At negative $x$, these solutions have the form
\begin{eqnarray}
  P_{\omega} = e^{- \pi 4 M \omega} e^{- i \omega \tau} e^{i 4 M \omega
  \tmop{Log} [- x]} & \label{P} &  \nonumber\\
  N_{\omega} = e^{+ \pi 4 M \omega} e^{- i \omega \tau} e^{i 4 M \omega
  \tmop{Log} [- x]} &  & \label{N} 
\end{eqnarray}
(for positive $x$ they of course agree with (\ref{shorizon})).

 Our purely outgoing mode $s_{\omega}$ can thus be decomposed as
\begin{equation}
  s_{\omega} = \frac{1}{e^{\pi 4 M \omega} - e^{- \pi 4 M \omega}} [e^{\pi 4 M
  \omega} P_{\omega} - e^{- \pi 4 M \omega} N_{\omega}], \label{sdecompose}
\end{equation}
and the negative free fall frequency part is then
\begin{equation}
  s_{\omega -} = \frac{e^{- \pi 4 M \omega}}{e^{- \pi 4 M \omega} - e^{\pi 4 M
  \omega}} N_{\omega} . \label{sminus}
\end{equation}
The complete wave-packet $s$ is then decomposed mode by mode to obtain the
negative free fall frequency piece $s_-$. \ To find the norm of $s_-$, we
introduce one additional wave packet, $s_{\tmop{reflect}}$:
$s_{\tmop{reflect}}$ is defined to be a reflection of $s$ around $x = 0$ at
very early times. \ Thus at early times it is bunched up against the
{\tmem{inside}} of the horizon, and vanishes {\tmem{outside}}. \ This
wave-packet corresponds to a particle which forms near the horizon, but then
falls inwards into the singularity. \ We similarly reflect the individual
modes (\ref{shorizon}). \ It is easy to check from the form (\ref{2dIP2}) of
the inner product that the norm of $s_{\tmop{reflect}}$ satisfies $<
s_{\tmop{reflect}}, s_{\tmop{reflect}} > = - < s, s >$, with an analogous
expression for the modes. \ \ \ It then follows from these definitions that
\begin{equation}
  N_{\omega} = s_{\omega} + e^{+ \pi 4 M \omega} s_{\tmop{reflect} \omega},
  \label{N2}
\end{equation}
which gives $< N_{\omega}, N_{\omega} > = < s_{\omega}, s_{\omega} > (1 -
e^{\pi 8 M \omega})$. \ We then use (\ref{sminus}) to obtain
\begin{equation}
  < s_-, s_- > = \frac{1}{1 - e^{\pi 8 M \omega}} < s, s > .
  \label{sminusnorm}
\end{equation}
Using (\ref{number}) and normalizing our wave-packets, this corresponds to a
thermal distribution of outgoing particles at the temperature $T = 1 / 8 \pi
M$.

\section{\label{GhostSection}Ghost Condensation}

The ghost condensate {\cite{ghost}} may be thought of as a degenerate case of
an irrotational perfect fluid. \ Indeed, if we consider an action of the form
\begin{equation}
  S_1 = \int \sqrt{- g} P (X) \label{Sone},
\end{equation}
with $X \equiv \nabla_{\mu} \varphi \nabla^{\mu} \varphi$ for some scalar
field $\varphi$, then the physical system which results is precisely
equivalent to that of an irrotational perfect fluid with pressure $P$ and
density $2 P' X - P$. \ Here $P$ is any generic function of $X$ of our
choosing. \

 The action (\ref{Sone}) yields an equation of motion
\begin{equation}
  2 P' (X)\Box \varphi + 4 P'' (X) \nabla^{\mu} \varphi \nabla^{\alpha}
  \varphi \nabla_{\mu} \nabla_{\alpha} \varphi = 0, \label{ghosteqnmot}
\end{equation}
and an energy-momentum tensor
\begin{equation}
  T^{\mu \nu} = 2 P' (X) \nabla^{\mu} \varphi \nabla^{\nu} \varphi - g^{\mu
  \nu} P (X) . \label{Tmunu}
\end{equation}
A ghost condensate background is defined to be a constant $X$ solution having
$P' = 0.$ \ By an appropriate choice of cosmological constant, we see from
(\ref{Tmunu}) that such solutions may be taken to have vanishing
energy-momentum. \ In flat space they correspond to Lorentz violating
backgrounds which we may take to have the form $\varphi_{} = A^2 t$, and if we
expand the field in small fluctuations around such backgrounds,
\begin{equation}
  \varphi = A^2 t + \pi, \label{pidef}
\end{equation}
the equation of motion at leading order becomes
\begin{equation}
  \text{$\ddot{\pi} = 0$.} \label{pidotdotzero}
\end{equation}
Note that the vanishing of $P'$ implies that the leading term in the
energy-momentum tensor will be linear in the $\pi$ fluctuations, and as a
result the theory violates the null energy condition. \ Now, generically we
should expect there to exist higher derivative operators in the action, and
these can stabilize the equation of motion (\ref{pidotdotzero}). \ We could
for example have a higher derivative term of the form
\begin{equation}
  S_2 = - \int \sqrt{- g}  \frac{2 \alpha \Box \varphi \Box
  \varphi}{\Lambda^2}, \label{Stwo}
\end{equation}
leading in flat space to the wave equation
\begin{equation}
  A^4 P'' (A^4) \ddot{\pi} + \alpha \frac{\Box^2 \pi}{\Lambda^2} = 0
  \label{wave},
\end{equation}
where $\Lambda$ is the UV cutoff.

At low energies, it follows from (\ref{wave}) that the spatial derivative
terms dominate over the time derivative terms, and we have
\begin{equation}
  A^4 P'' (A^4) \ddot{\pi} + \alpha \frac{\overrightarrow{\nabla^{}}^4
  \pi}{\Lambda^2} = 0. \label{lowewave}
\end{equation}
In fact, this low energy form of the wave equation in flat space is generic
and does not depend on a specific assumption for the higher derivative
operators such as (\ref{Stwo}).

 To keep expressions simpler we will henceforth adopt a specific
form for $P (X)$:
\begin{equation}
  P (X) = \frac{1}{2 \Lambda^4} (X - \Lambda^4)^2 . \label{PX}
\end{equation}
This choice does not influence the results since we are interested only in the
general form of the wave equation. \ The ghost condensate background in flat
space is then
\begin{equation}
  \varphi = \Lambda^2 t, \label{flatbackground}
\end{equation}
and the low energy dispersion relation following from (\ref{lowewave}) is
\begin{equation}
  \omega^2 = \alpha \frac{k^4}{\Lambda^2} . \label{dispersion}
\end{equation}
 We next consider the ghost condensate theory in the presence of a
Schwarzschild black hole in 3+1 dimensions. The metric again takes the form
(3), but now with the addition of the usual angular coordinates:
\begin{equation}
  \tmop{ds}^2 = \left( 1 - v^2 \right) \mathd \tau^2 + 2 v \tmop{drd} \tau -
  \tmop{dr}^2 - r^2 d \Omega^2 . \label{metric}
\end{equation}
In the presence of the black hole, solutions of constant $X$ having $P' = 0$
continue to solve the equation of motion (\ref{ghosteqnmot}). \ In this case
$\nabla^{\mu} \varphi$ now follows the world-lines of observers who freely
fall into the black hole from rest at infinity {\cite{shinji}}. \ This
corresponds to a background of the form
\begin{equation}
  \varphi_B = \Lambda^2 \tau \label{background} .
\end{equation}
Note that unlike in flat space, the higher derivative terms in the equation of
motion will perturb this solution and in fact lead to a gradual accretion of
ghost condensate energy into the black hole. \ These effects are small however
if we take
\begin{equation}
  \frac{1}{2 M} \ll \frac{\Lambda}{\sqrt{\alpha}} \label{approx1}
\end{equation}
and
\begin{equation}
  \Lambda \ll M_{\tmop{pl}}, \label{approx2}
\end{equation}
which we will assume throughout the remainder of the paper. \ The first
inequality is just the statement that the Schwarzschild radius of the black
hole is much larger than the short distance cutoff of the effective field
theory, which is certainly required for consistency. \ Having $\Lambda$ much
smaller than the $\tmop{Planck}$ scale suppresses the Jeans instability in the
ghost condensate, and also suppresses mixing between the $\pi$ excitations and
the graviton.

 We may now expand the equation of motion (\ref{ghosteqnmot}) about
the background (\ref{background}) to linear order in the $\pi$ fluctuations. \
We obtain
\begin{equation}
  \nabla_{\nu} \xi^{\nu} \xi^{\mu} \nabla_{\mu} \pi + \xi^{\mu} \xi^{\nu}
  \nabla_{\mu} \nabla_{\nu} \pi + 2 \xi^{\mu} \nabla_{\mu} \xi^{\nu}
  \nabla^{}_{\nu} \pi = 0, \label{pieqnmot}
\end{equation}
where we have defined $\xi^{\mu} \equiv \frac{1}{\Lambda^2} \nabla^{\mu}
\varphi_B$ to simplify expressions. \ As noted above, $\xi^{\mu}$ follows the
4-velocities of observers falling freely into the black hole along geodesics,
and as a result the third term in (\ref{pieqnmot}) vanishes by the geodesic
equation. \ This leaves us with
\begin{equation}
  \nabla_{\nu} \xi^{\nu} \xi^{\mu} \nabla_{\mu} \pi + \xi^{\mu} \xi^{\nu}
  \nabla_{\mu} \nabla_{\nu} \pi = 0. \label{pieqnmot2}
\end{equation}
This wave equation follows from an action for $\pi$ of the form
\begin{equation}
  S_1^{\pi} = \int \sqrt{- g} (\xi^{\mu} \nabla_{\mu} \pi)^2 .
  \label{piaction}
\end{equation}
Now, we also have to include a higher derivative term in this action in order
to stabilize the fluctuations. \ In the next section we turn to a discussion
of the choice of this higher derivative term.

\section{\label{higher}Higher Derivative Operators}

As pointed out in section (\ref{GhostSection}) the choice of higher derivative
operators typically does not affect the qualitative features of the low energy
effective field theory of the ghost condensate particles. \ Unfortunately,
just about all calculations performed to date of Hawking spectra, including
the one reviewed in section (\ref{2D}), make references to arbitrarily short
distance physics. \ The detailed calculation presented here will also have
this unpleasant feature, and we will be having to choose some particular
higher derivative terms to include in our action.

 Fortunately, there is good reason to think that the results of the
usual calculations, as well as this one, are independent of most of the
assumptions made in the UV {\cite{nice}}. \ In particular, one can imagine a
calculation scheme to make the UV independence manifest, in the following way:

 Consider foliating Schwarzschild space-time into a family of
space-like hyper-surfaces- ``nice slices''- which have the property that their
extrinsic curvature is never much greater than $\sim 1 / M$. \ We also assume
that at infinity the time directions normal to the nice slices agree with the
time-like killing vectors of the asymptotically flat space-time. \ That a
family of nice slices actually exists was demonstrated in {\cite{nice}}.

 Now, we can imagine following the evolution of the state of the
black hole from before it was formed, to some very late time afterwards. \
Hawking radiation would appear as low energy outgoing particles at large
distances from the hole. \ The time evolution in question will be generated by
a time dependent Hamiltonian operator, since the time direction normal to the
nice slices is generally not the killing direction. \ On the other hand, all
length scales governing the time evolution are at least of order the
Schwarzschild radius. \ It then follows from the adiabatic theorem that modes
of energy much larger than $\sim 1 / M$ will not be excited if they started in
their ground states.

 Now, in the case of a theory with spontaneous Lorentz violation we
have to be careful, since the modes we have argued are unexcited have high
energy {\tmem{as viewed by observers moving normal to the nice slices}}. \
What matters for the validity of the low energy effective theory of the ghost
condensate, however, is that there be no modes excited which have large energy
with respect to the direction picked out by $\nabla^{\mu} \varphi_B$. \ The
key point, though, is that slices of constant $\varphi_B$ have all the
required properties to be nice slices {\tmem{so long as we do not look deep
inside the horizon}}.

 The extrinsic curvature on a surface of constant $\varphi_B$ is
given by{\footnote{See Wald ``General Relativity'' page 230 for a
definition.}}
\begin{equation}
  K = \frac{\Box \varphi_B}{\sqrt{X}} = - \frac{3}{2 r}  \sqrt{\frac{2 M}{r}}
  \label{curvature} .
\end{equation}
Here $\Box \varphi_B$ was evaluated using the Christoffel symbols for the
metric (\ref{metric}) which may be found in appendix A. \ This shows that as
long as one is only interested in processes occurring outside of the horizon
at $r = 2 M$, then surfaces of constant $\varphi_B$ have acceptable extrinsic
curvature to serve as nice slices. \ That these slices also have the correct
asymptotic behavior at infinity follows from expressions (\ref{tau}) and
(\ref{background}).

 It thus follows that low energy effective field theory can be
expected to be valid for the calculation of the Hawking quanta emitted by a
black hole in the ghost condensate.{\footnote{Though not perhaps for
determining the quantum state after evaporation as was argued for in the
standard case using nice slices.}} \ For this argument we do however have to
assume that whatever the true UV completion of the ghost condensate may be (if
one exists), it does not allow signal propagation at speeds much greater than
the speed of light. \ Similarly, for the purposes of our calculation using the
$\pi$ wave equation, we will require that the higher derivative terms we add
to the action leave us with a well defined light cone for signal propagation.

 Perhaps surprisingly, it is actually rather hard to find higher
derivative terms which have this property. \ In flat space, if we write a wave
equation as
\begin{equation}
  F (i \partial_t, i \partial_x) \pi = 0, \label{F}
\end{equation}
then it turns out that the requirement for strictly subluminal propagation is
for $F (\omega, p)^{- 1}$ to be analytic in the region $\tmop{Im} \omega > |
\tmop{Im} p|$ {\cite{cone}}. \ A simple class of higher derivative terms which
leave the $\pi$ wave equation with the appropriate analytic property may be
added to (\ref{piaction}) to yield the family{\footnote{Note that we expect
the Hawking spectrum to be well approximated by a calculation in the free
field theory of the ghost phonons so long as interactions may be neglected. \
This is valid so long as energies of excited modes are much below the cutoff,
implying that evolution due to the nonrenormalizable interactions is
suppressed.}}
\begin{equation}
  S^{\pi} = S^{\pi}_1 - \int \sqrt{- g}  \frac{\alpha [(\Box+ ( \frac{1}{c^2}
  - 1) \xi^{\mu} \xi^{\nu} \nabla_{\mu} \nabla_{\nu}) \pi]^2}{\Lambda^2},
  \label{piaction2}
\end{equation}
where $\alpha$ and $c$ are free parameters which we will assume to be of order
one.

 In flat space (\ref{piaction2}) yields the dispersion relation
\begin{equation}
  \omega^2 = \frac{\alpha}{\Lambda^2} (k^2 - \frac{\omega^2}{c^2})^2
  \label{generaldisperse} .
\end{equation}
Note that at energies far below $\Lambda$ there are modes satisfying the usual
ghost dispersion relation (\ref{dispersion}). \ At energies far above
$\Lambda$ however, the higher derivative terms in the wave equation dominate,
and signals propagate at the speed $c$.{\footnote{There are however negative
energy states in the spectrum which we will have to treat carefully.}} \ It is
nice that the parameter $c$ gives us one check on possible UV sensitivity of
our results- \ in fact we will see that the value of $c$ does not affect the
Hawking spectrum in ghost phonons to within the approximations we employ.

 It should be pointed out that, while we expect the Hawking
spectrum to be independent of the UV physics, there is an additional subtlety:
\ Although in flat space-time, the low energy wave equation satisfied by the
$\pi$'s does not depend on the choice of higher derivative terms, this is not
generally true in curved space-time. \ At this stage it is conceivable that
different higher derivative terms, leading to different {\tmem{low energy}}
physics in curved space-time could lead to different results for the Hawking
spectrum. \ In the next section however, we will give some general arguments
that this is not the case.

 Finally we note that we have dropped a source term in the wave
equation which arises due to the fact that in the presence of the higher
derivative operators the background (\ref{background}) is not an exact
solution to the equation of motion. \ The true background is a small
perturbation to this one under the approximation (\ref{approx1}) that the
black hole is large, and thus the wave equation we are using is also valid in
this regime.

\section{\label{spectrum}The Hawking Spectrum of Ghost Phonons}

Equipped with the action (\ref{piaction2}), and the metric (\ref{metric}), we
now set out to calculate the Hawking spectrum in $\pi$ particles, following
the same line of argument presented for the 2D massless scalar in section
(\ref{2D}). \ For simplicity we will restrict ourselves to calculating the
flux in s-wave modes and so will ignore $\pi$ dependence on the angular
variables.

 In the ($\tau, r$) coordinate system the wave equation following
from (\ref{piaction2}) takes the form
\begin{eqnarray}
  \left[ \partial_{\tau}^2 + 2 v \partial_{\tau} \partial_r + v^2 \partial_r^2
  - \frac{v}{2 r} \partial_{\tau} - \frac{v^2}{r} \partial_r +
  \frac{v^2}{r^2}
 +\frac{\alpha}{\Lambda^2} \left( \frac{1}{c^2} \partial_{\tau}^2
  + \frac{2 v}{c^2} \partial_{\tau} \partial_r - \right. \right. \nonumber\\
\left. (1 - \frac{v^2}{c^2})
  \partial_r^2  +  ( \frac{3 v}{r} - \frac{2 v}{c^2 r}) \partial_{\tau} + (
  \frac{3 v^2}{2 r} - \frac{5 v^2}{2 c^2 r}) \partial_r + (- \frac{7 v^2}{2
  r^2} + \frac{9 v^2}{2 c^2 r^2}) \right)^2 + \nonumber\\ \frac{\alpha}{\Lambda^2} (
  \frac{1}{c^2} - 1) ( \frac{3 v}{r}) (\partial_{\tau} + v \partial_r +
  \frac{v}{2 r}) \left( \frac{1}{c^2} \partial_{\tau}^2 + \frac{2 v}{c^2}
  \partial_{\tau} \partial_r - (1 - \frac{v^2}{c^2}) \partial_r^2 + \right. \nonumber
  \\\left. \left.( \frac{3
  v}{r} - \frac{2 v}{c^2 r}) \partial_{\tau} + ( \frac{3 v^2}{2 r} - \frac{5
  v^2}{2 c^2 r}) \partial_r + (- \frac{7 v^2}{2 r^2} + \frac{9 v^2}{2 c^2
  r^2})\right) \right] \tilde{\pi} = 0, \label{mess}
\end{eqnarray}
where we have made the usual change of variables in 4D from $\pi$ to $\pi =
\frac{\tilde{\pi}}{r}$. \ Although this expression certainly looks like a
mess, the reader will be glad to know that it is about to simplify
significantly. \

 As in section (\ref{2D}) we are interested in solutions to this
equation corresponding to modes which are purely outgoing at infinity and
which vanish behind the horizon (note that depending on the value of $c$ the
sound horizon for the phonons actually is located at different radii, and it
coincides with the usual black hole horizon for c = 1). \ Again we will look
for solutions of the form
\begin{equation}
  \tilde{\pi} = e^{- i \omega \tau} e^{i \int^r k (r') \tmop{dr}'}
  \label{ansatz}
\end{equation}
and hope that a WKB approximation will be valid (c.f. (\ref{WKB1}) and
(\ref{WKB2})). \ The WKB approximation allows us to drop terms in (\ref{mess})
in which an r-derivative acts on $v$ instead of on $\tilde{\pi}$. \
Equivalently, since $\frac{\partial v}{\partial r} \sim \frac{v}{r}$, we may
also drop terms in which we have a factor of $\frac{1}{r}$ instead of an
r-derivative on $\widetilde{\pi}$. \ The fact that this approximation is
indeed valid will follow once we have obtained an expression for $k (r)$. \
Its validity will turn out to be equivalent to the requirement $\frac{1}{2 M}
\ll \frac{\Lambda}{\sqrt{\alpha}}$ that the black hole is much larger than the
short distance cutoff. \ With the sub-dominant terms dropped we obtain the
much simplified wave equation
\begin{equation}
  \left[ \partial_{\tau}^2 + 2 v \partial_{\tau} \partial_r + v^2 \partial_r^2
  + \frac{\alpha}{\Lambda^2} \left( \frac{1}{c^2} \partial_{\tau}^2 + \frac{2
  v}{c^2} \partial_{\tau} \partial_r - (1 - \frac{v^2}{c^2}) \partial_r^2
  \right)^2 \right] \tilde{\pi} = 0. \label{lessmess}
\end{equation}
With the ansatz (\ref{ansatz}) and the second piece of the WKB approximation
(\ref{WKB2}), this becomes
\begin{equation}
  (\omega - v k)^2 = \frac{\alpha}{\Lambda^2} (k^2 - \frac{1}{c^2} (\omega - v
  k)^2)^2 \label{curvedisperse} .
\end{equation}
Solving for $k (r)$ we obtain:
\begin{equation}
  k (r) = \frac{- (\pm 1 + \frac{\sqrt{\alpha}}{\Lambda} \frac{2}{c^2} \omega)
  v \pm' \sqrt{v^2 (\pm 1 + \frac{\sqrt{\alpha}}{\Lambda} \frac{2}{c^2}
  \omega)^2 + 4 \frac{\sqrt{\alpha}}{\Lambda} (1 - \frac{v^2}{c^2}) (\pm
  \omega + \frac{\sqrt{\alpha}}{\Lambda} \frac{\omega^2}{c^2})}}{2
  \frac{\sqrt{\alpha}}{\Lambda} (1 - \frac{v^2}{c^2})} \label{kmess},
\end{equation}
where the prime on $\pm'$ indicates that it is separate from the other
instances of $\pm$ (there are 4 solutions to (\ref{curvedisperse})). \ Now
note that $\omega$ is the frequency of the mode as seen by an observer at
infinity, so that we may take $\omega \ll \Lambda$. \ In addition, having a
mode with positive frequency $\omega$ and real wave number corresponds to a
specific choice of the signs, so that the expression for $k$ becomes
\begin{equation}
  k (r) = \frac{- v + \sqrt{v^2 + 4 \frac{\sqrt{\alpha}}{\Lambda} (1 -
  \frac{v^2}{c^2}) \omega}}{2 \frac{\sqrt{\alpha}}{\Lambda} (1 -
  \frac{v^2}{c^2})} \label{ksimple} .
\end{equation}
From this expression it is a straightforward exercise to confirm that the WKB
constraints (\ref{WKB1}) and (\ref{WKB2}) are indeed satisfied.

 In fact our expression for $k (r)$ already reveals the physical
reason for the absence of the Hawking effect in this theory. \ First of all,
if we consider the limit of very small $\omega$, then we can see that $k$
becomes of order the cutoff, $\Lambda$, when $v$ becomes of order 1 (again,
this is assuming that $c$ and $\alpha$ are also order 1). \ That is, $k \sim
\Lambda$ for $r$ of order {\tmem{but not necessarily close to}} the horizon
radius. \ For larger $\omega$ this will occur at still farther radii. \ This
is to be contrasted with the standard case considered in section (\ref{2D}),
for which such substantial blue shifting of the mode doesn't occur until it is
located perturbatively close to the horizon. \ Along the same lines, let us
choose an energy $E$ satisfying $\Lambda \gg E \gg 1 / M$. \ Then it follows
from (\ref{wff}) and (\ref{ksimple}) that $\omega_{\tmop{ff}} \sim E$ when
$\frac{r}{2 M} \sim \frac{\Lambda}{E}$. \ Note now that this is a result which
does not require pushing the limits of the effective field theory since it
only involves energies much smaller than $\Lambda$. \ We thus see that the
outgoing modes have energies much larger than $1 / M$ very far away from the
horizon, while simultaneously satisfying the WKB approximation. \ It follows
that the negative frequency components of these packets are negligible; by the
WKB approximation they appear just like packets constructed out of ordinary
plane waves at the given radius, and with the appropriate free fall frequency.
\ Since these outgoing packets are then composed of only positive frequency
components with energies much larger than $1 / M$, we may conclude that (to
within the approximations employed) ghost condensate quanta are not radiated
by these black holes.{\footnote{ \ Making this argument more precise would
involve going through an analysis with wave-packets such as that found in
{\cite{packetradiation}}.}} \ Again, this argument is independent of
assumptions about the behavior of the theory at high energies above the
cutoff, and \ in addition, it is also independent of the low energy effects of
the choice of higher dimension operators: \ In the WKB approximation, the
wave-packets propagate locally as if they were in flat space, and in flat
space, the low energy behavior of the modes is generic, as explained in
section (\ref{GhostSection}).

 On the other hand, the reader may be wondering if one could
actually follow through with the same sort of near-horizon, infinite-blue
shift argument given in section (\ref{2D}) for the 2D massless scalar. \ We
will now demonstrate that this is in fact possible, and that it yields a
confirmation of our result.

 As in section (\ref{2D}), the action (\ref{piaction2}) leads to a
conserved inner product on complex solutions to the wave equation. \ This
inner product is given by{\footnote{The appropriate inner product may be
determined as follows: \ Our action is a bilinear in $\pi$ with a sum of terms
of the form $\int D_{\Alpha} \pi D_B \pi$ for some differential operators
$D_A$ and $D_B$. \ Now consider a new action given by replacing each of these
terms by $\int D_A s_1^{\ast} D_B s_2$. This new action has a $U (1)$ symmetry
under which $s_1$ and $s_2$ transform with the same charge. \ Let $J^{\mu}$ be
the associated conserved current. \ Then the inner product is given by
$\int_{\Sigma} n^{\mu} J_{\mu}$, with the overall coefficient chosen to
enforce the relation (\ref{aalgebra}), with an appropriate normalization for
the creation/annihilation operators.}}
\begin{eqnarray}
  < s_1, s_2 > &\equiv& i \int_{\Sigma} n^{\mu} \left[ \xi_{\mu} s_1^{\ast} \xi^{\nu}
  \nabla_{\nu} s_2 - \frac{\alpha}{\Lambda^2} \left(\nabla_{\mu} s_1^{\ast} + (
  \frac{1}{c^2} - 1) \xi_{\mu} \xi^{\nu} \nabla_{\nu}
  s_1^{\ast}\right) \bigg(\Box s_2 \right. \nonumber\\ &+& \left. (
  \frac{1}{c^2} - 1) \xi^{\alpha} \xi^{\beta} \nabla_{\alpha} \nabla_{\beta}
  s_2\right) + \frac{\alpha}{\Lambda^2} s_1^{\ast} \left(\nabla_{\mu} \Box s_2 + (
  \frac{1}{c^2} - 1) (\xi_{\mu} \xi^{\nu} \nabla_{\nu} \Box s_2  \right. \nonumber\\
  &+& 2 \nabla_{\mu} \xi^{\alpha} \xi^{\beta} \nabla_{\alpha} \nabla_{\beta} s_2 +
  \xi^{\alpha} \xi^{\beta} \nabla_{\alpha} \nabla_{\beta} \nabla_{\mu} s_2 +
  \nabla_{\alpha} \xi^{\alpha} \xi_{\mu} \Box s_2)  \nonumber\\ &+&\left.( \frac{1}{c^2} - 1)^2
  (\xi_{\mu} \xi^{\nu} \xi^{\alpha} \xi^{\beta} \nabla_{\nu} \nabla_{\alpha}
  \nabla_{\beta} s_2 + \nabla_{\nu} \xi^{\nu} \xi_{\mu} \xi^{\alpha}
  \xi^{\beta} \nabla_{\alpha} \nabla_{\beta} s_2)\right) \nonumber\\  &-& (s_1^{\ast}
  \leftrightarrow s_2)\bigg]. \label{IPmess}
\end{eqnarray}
On a surface of constant $\tau$ the normal vector is $n^{\mu} = \xi^{\mu}$,
and the expression simplifies to
\begin{eqnarray}
  < s_1, s_2 >&=& i \int_{\tau = \tmop{const}} \left[s_1^{\ast} \xi^{\nu}
  \nabla_{\nu} s_2 - \frac{\alpha}{\Lambda^2} ( \frac{1}{c^2} \xi^{\nu}
  \nabla_{\nu} s_1^{\ast}) \left(\Box s_2 + ( \frac{1}{c^2} - 1) \xi^{\alpha}
  \xi^{\beta} \nabla_{\alpha} \nabla_{\beta} s_2\right) \right. \nonumber\\ &+& \frac{\alpha}{\Lambda^2}
  s_1^{\ast} \left( \frac{1}{c^2} \xi^{\mu} \nabla_{\mu} \Box s_2 + ( \frac{1}{c^2}
  - 1) (( \frac{1}{c^2} - 1) \nabla_{\nu} \xi^{\nu} \xi^{\alpha} \xi^{\beta}
  \nabla_{\alpha} \nabla_{\beta} s_2 \right.\nonumber\\ &+& \left. \left.\frac{1}{c^2} \xi^{\nu} \xi^{\alpha}
  \xi^{\beta} \nabla_{\nu} \nabla_{\alpha} \nabla_{\beta} s_2 + \nabla_{\nu}
  \xi^{\nu} \Box s_2)\right) - (s_1^{\ast} \leftrightarrow s_2)\right] . \label{IPlessmess}
\end{eqnarray}
We may use this inner product as before to associate annihilation operators
with solutions to the wave equation of positive norm.{\footnote{We may still
use expression (\ref{a}) to define $a (s)$, and the canonical commutation
relations will still yield (\ref{aalgebra}). \ Note however that one must be
careful to use the correct canonical commutation relations appropriate to a
higher time-derivative field theory. \ See e.g. {\cite{Weldon}}.}} \ We are
again interested in evaluating $< a^{\dag} (s) a (s) >$ for a purely outgoing
wave-packet $s$, and will do so by evaluating the inner product defining $a
(s)$ at an early time when the wave packet is bunched up against the horizon
(for $c \neq 1$, against the sound horizon). \ We can then decompose the
wave-packet into a basis of high free fall frequency plane waves, enabling us
to apply the condition that our state contains no high energy excitations.

 It is important to notice, however, that due to the fact that our
action contains higher derivatives with respect to time, there is a
complication. \ At high energies above $\Lambda$ there will be ghost modes
present, and, being negative energy states, these modes have the opposite
relationship between their norm in the inner product (\ref{IPlessmess}) and
the sign of their free fall frequency.{\footnote{Unfortunately in order to use
standard terminology, we are now using the term ``ghost'' to refer to
negative energy states, not to general perturbations of the ghost 
condensate background. Please see {\cite{Linde}} for a discussion of the quantization of
this theory.}} \ We note that the presence of these ghost modes in our
calculation is not a problem since they occur only above the high scale
$\Lambda$, and moreover we are now working with a free field theory- the
negative energy states do not lead to an instability. \ On the other hand,
when we decompose our wave-packet into pieces of positive and negative norm,
it is no longer sufficient to decompose into pieces of positive and negative
free fall frequency. \ The positive norm part of the wave-packet will now
include a {\tmem{negative frequency ghost part}}. \ In fact, we will now show
that the negative norm part of the wave-packet is entirely absent near the
horizon- the negative frequency piece we would usually associate with particle
creation is actually composed entirely of ghost modes and thus has
{\tmem{positive norm}}. \ Likewise, the positive frequency piece is actually
free of the ghost modes. \ It is in this way that we will confirm our earlier
analysis which showed an absence of phonon Hawking radiation in this model.

 Near the sound horizon at $r = \frac{2 M}{c^2}$, the full
expression (\ref{kmess}) for the wavenumber (with appropriate sign choices)
becomes at leading order
\begin{equation}
  k (x) \simeq \frac{2 M \Lambda}{c \sqrt{\alpha}} \label{kapprox} (1 +
  \frac{2 \sqrt{\alpha} \omega}{c^2 \Lambda}) \frac{1}{x},
\end{equation}
where $x = r - \frac{2 M}{c^2}$. \ The expression for our outgoing mode
outside the horizon then takes the form
\begin{equation}
  \tilde{s}_{\omega} \simeq  e^{- i \omega \tau} e^{i
  \frac{2 M \Lambda}{c \sqrt{\alpha}} (1 + \frac{2 \sqrt{\alpha} \omega}{c^2
  \Lambda}) \tmop{Log} [x]} . \label{modeapprox}
\end{equation}
Positive and negative free fall frequency modes can be obtained by
analytically continuing in the upper or lower half complex $x$ plane, as
before. \ At negative $x$ these modes take the form
\begin{equation}
  \tilde{P}_{\omega} = e^{- \pi \frac{2 M \Lambda}{c \sqrt{\alpha}}} e^{- i
  \omega \tau} e^{i \frac{2 M \Lambda}{c \sqrt{\alpha}} \tmop{Log} [- x]}
  \label{Pghost}
\end{equation}
\begin{equation}
  \tilde{N}_{\omega} = e^{+ \pi \frac{2 M \Lambda}{c \sqrt{\alpha}}} e^{- i
  \omega \tau} e^{i \frac{2 M \Lambda}{c \sqrt{\alpha}} \tmop{Log} [- x]},
  \label{Nghost}
\end{equation}
where we have now dropped the sub-leading terms in $\frac{\omega}{\Lambda}$.

 Again, a purely outgoing wave-packet which vanishes behind the
horizon can be formed as a linear combination of modes of these types. \ We
will now show, however, that when we decompose $P_{\omega}$ and $N_{\omega}$
into definite free fall frequency plane waves near the horizon, $P_{\omega}$
will be composed purely of regular plane waves, while $N_{\omega}$ will be
composed purely of ``ghost'' plane waves.

 In flat space the dispersion relation (\ref{curvedisperse})
becomes
\begin{equation}
  \omega^2 = \frac{\alpha}{\Lambda^2} (k^2 - \frac{\omega^2}{c^2})^2 .
  \label{flatdisperse}
\end{equation}
This has four solutions:
\begin{equation}
  k = \pm \sqrt{\frac{w^2}{c^2} \pm'  \frac{\Lambda}{\sqrt{\alpha}} \omega}
  \label{flatk} .
\end{equation}
The ghost modes correspond to choosing the $-$ sign inside the square root. \
At low energies these modes have imaginary $k$, but for $\omega \gtrsim
\Lambda$, they become regular plane waves, despite having negative norm in the
inner product (\ref{IPmess}). \ If we take $w \gg \Lambda$, we can approximate
the modes as having
\begin{equation}
  k \sim \pm \frac{\omega}{c} \pm (\pm'  \frac{c \Lambda}{2 \sqrt{\alpha}}) .
  \label{kghost} 
\end{equation}
 Now, the modes we will expand our wave-packet in will be high
frequency plane waves in the vicinity of the horizon as seen by a free falling
observer, having the form
\begin{equation}
  \tilde{\pi} \sim e^{- i \omega \Delta \tau} e^{i k \Delta R}
  \label{planewaves},
\end{equation}
where R is a spatial coordinate orthogonal to $\tau$ (and the angular
directions) near the horizon. \ From the metric (\ref{metric}) we can write $R
\simeq x + c \tau$, so that the plane waves of interest become
\begin{equation}
  \tilde{\pi} \sim e^{- i (\omega - c k) \Delta \tau} e^{i \tmop{kx}} .
  \label{planewaves2}
\end{equation}
We will use this form for the plane waves to explicitly calculate the
components of $s$. \ Before we do so however, we will first give an argument
which will show the same results with less work. \ For simplicity we will here
consider the case $c = 1$.

 Along outgoing null rays with $\Delta R \sim \Delta \tau$, the
positive frequency plane waves (\ref{planewaves}) have leading
behavior  \
\begin{eqnarray}
  \tilde{\pi}_{\tmop{rg}} & \sim & e^{- i \frac{\Lambda}{2 \sqrt{\alpha}}
  \Delta \tau} \nonumber\\
  \tilde{\pi}_{\lg}  & \sim & e^{- 2 i \omega \tau} e^{i \frac{\Lambda}{2
  \sqrt{\alpha}} \Delta \tau} \nonumber\\
  \tilde{\pi}_{\tmop{rn}}  & \sim & e^{i \frac{\Lambda}{2 \sqrt{\alpha}}
  \Delta \tau} \nonumber\\
  \tilde{\pi}_{\ln} & \sim & e^{- 2 i \omega \tau} e^{- i \frac{\Lambda}{2
  \sqrt{\alpha}} \Delta \tau}.  \label{fourmodes}
\end{eqnarray}
Here the letters $r$ and $l$ label right going or left going modes, while the
letters $n$ and $g$ label normal or ghost type modes. \ Now note that it
follows from the metric (\ref{metric}) that outgoing null rays near the
horizon actually follow paths of the form $x \sim A (1 + \frac{\Delta \tau}{4
M})$, where $\Delta \tau$ and the constant $A$ here are both taken much
smaller than the Schwarzschild radius. \ Along such paths the positive
frequency mode $\tilde{P}_{\omega}$, from (\ref{Pghost}), has the leading
behavior $\tilde{P}_{\omega} \sim e^{i \frac{\Lambda}{2 \sqrt{\alpha}} \Delta
\tau}$. \ It follows that, when we decompose $\tilde{P}_{\omega}$ into the
four types of modes in (\ref{fourmodes}), the left going modes cannot be
present, and neither can the right going ghost modes. \ Only a superposition
of the right going normal type modes will yield an overall factor of $e^{i
\frac{\Lambda}{2 \sqrt{\alpha}} \Delta \tau}$.{\footnote{One could imagine
cancelling an incorrect $\tau$ dependence along a null ray by using
combinations of plane waves with different values of $\omega$. \ This
possibility is excluded, however, since we may choose any null ray we like; \
such a cancellation could not be made to work across all outgoing null rays
simultaneously.}} \ We thus conclude that $\tilde{P}_{\omega}$ consists
entirely of normal type modes.

 The interesting point now is that an analogous argument reveals
that $\tilde{N}_{\omega}$ is composed entirely of ghost type modes. \ The
reason is that the plane waves with {\tmem{negative}} free fall frequency may
be obtained from their positive frequency counterparts (\ref{fourmodes}) by
complex conjugation. \ In particular, the negative frequency, right going
{\tmem{ghost}} modes will have the behavior \ $e^{i \frac{\Lambda}{2
\sqrt{\alpha}} \Delta \tau}$ along outgoing null rays. \ As advertised, we
then conclude that the negative frequency part of the outgoing mode $s$ is
composed entirely of ghosts, and thus has {\tmem{positive}} norm, in contrast
to the standard case. \ It follows that the Hawking effect is absent, at least
to within the approximations we have been employing.

 We now turn to an explicit calculation of the components of $s$
using the expression (\ref{IPlessmess}) for the inner product. \ We will use
the fact that $\frac{1}{2 M} \ll \Lambda$, as well as the WKB approximation. \
Note that the WKB approximation is valid for the $\tilde{s}_{\omega}$, as well
as for the plane wave modes (\ref{planewaves2}) at high frequencies. \ The
inner product can then be written{\footnote{More precisely, to justify the
near horizon approximation we are employing we should use the full
wave-packets formed out of the various modes. \ This can be done
straightforwardly and does not affect the analysis.}}
\begin{eqnarray}
  < s_w, s_2 > &=& 4 \pi i \int_{r > 2 m} \tmop{dr} [ \tilde{s}_{\omega}^{\ast}
  (\partial_{\tau} + v \partial_r) \tilde{s}_2 - \frac{\alpha}{\Lambda^2 c^2}
  (\partial_{\tau} + v \partial_r) \widetilde{s_{\omega}}^{\ast} (
  \frac{1}{c^2} (\partial_{\tau} + v \partial_r)^2 - \partial_r^2) \tilde{s}_2
  \nonumber\\ &+& \frac{\alpha}{\Lambda^2 c^2} \tilde{s}_{\omega}^{\ast} (\partial_{\tau} +
  v \partial_r) ( \frac{1}{c^2} (\partial_{\tau} + v \partial_r)^2 -
  \partial_r^2) \tilde{s}_2 - ( \widetilde{s_{\omega}}^{\ast} \leftrightarrow
  \tilde{s}_2)], \label{IPsimple}
\end{eqnarray}
where as usual the tilde indicates that we have multiplied by r, and we only
integrate outside the horizon since $s_{\omega}$ vanishes inside. \ We will
take $s_2$ to be one of the high frequency plane wave modes
(\ref{planewaves2}). \ In evaluating the integral we will make use of the
identities
\begin{eqnarray}
  ( \frac{1}{c^2} (\partial_{\tau} + v \partial_r)^2 - \partial_r^2)
  \tilde{s}_{\omega}^{\ast} & = & - i \frac{\Lambda}{\sqrt{\alpha}} v
  \partial_r \tilde{s}_{\omega}^{\ast} \nonumber\\
  ( \frac{1}{c^2} (\partial_{\tau} + v \partial_r)^2 - \partial_r^2)
  \tilde{s}_2 & = & \pm' \frac{\Lambda}{\sqrt{\alpha}} \omega_2  \tilde{s}_2 .
  \label{identities}
\end{eqnarray}
In these expressions, $\omega_2$ is the frequency of $s_2$ appearing in
expression (\ref{planewaves2}), and the $\pm'$ coincides with that of
expression (\ref{flatk}). \ In the first identity, we have dropped a
sub-dominant term proportional to $\omega$. \ The largest term that we can get
from the integral (\ref{IPsimple}) will be proportional to $\omega_2^2$, and
we will drop all terms with weaker $\omega_2$ dependence. \ This is an
approximation which can be made arbitrarily accurate by computing the integral
at earlier and earlier times; as the outgoing wave-packet is compressed
against the horizon it becomes composed of arbitrarily high $\omega_2$
oscillations. \ A cutoff on $\omega_2$ for plane waves with appreciable
overlap with $s_{\omega}$ would appear if we actually formed a wave-packet to
localize the integral slightly away from the horizon. \ At earlier and earlier
times, however, this cutoff becomes arbitrarily high. \ This observation
allows us in particular to \ immediately discard the single derivative terms
in (\ref{IPsimple}). \ Similarly, we will keep only leading terms in $x$.
Putting all of this together, the inner product can be written
\begin{equation}
  < s_{\omega}, s_2 > = 4 \pi i \frac{\sqrt{\alpha}}{\Lambda c^2} \int_{r > 2
  m} \mp' i \omega_2^2 \tilde{s}_{\omega}^{\ast} \tilde{s}_2 + (1 \pm' 1) c
  \omega_2 \tilde{s}_2 \partial_r \widetilde{s_{\omega}}^{\ast} + \tmop{ic}^2
  \tilde{s}_2 \partial^2_r \widetilde{s_{\omega}}^{\ast} . \label{IPsimple2}
\end{equation}

We will now integrate by parts to put all of the r derivatives on
$\tilde{s}_2$. \ We can then use $\partial_r \tilde{s}_2 = i k_2  \tilde{s}_2$
to obtain
\begin{equation}
  < s_{\omega}, s_2 > = 4 \pi \frac{\sqrt{\alpha}}{\Lambda c^2} \int_{r > 2 m}
  (\pm' \omega_2^2 + k_2 (1 \pm' 1) c \omega_2 + k_2^2 c^2)
  \widetilde{s_{\omega}}^{\ast} \tilde{s}_2 . \label{IPsimple3}
\end{equation}
With expression (\ref{kghost}) for the possible values of $k_2$, this becomes
at leading order
\begin{equation}
  < s_{\omega}, s_2 > = 4 \pi \frac{\sqrt{\alpha}}{\Lambda c^2} \int_{r > 2 m}
  (\pm' \omega_2^2 \pm (1 \pm' 1) \omega^2_2 + \omega^2_2)
  \widetilde{s_{\omega}}^{\ast} \tilde{s}_2 . \label{IPsimple4}
\end{equation}

It then follows that the order $\omega_2^2$ contribution to the inner product
is non-zero only when $s_2$ is either a positive frequency, right going,
normal type mode, or a negative frequency, right going, ghost type mode. \ As
argued above, this again allows us to reach the conclusion that, to within the
approximations employed, \ phonon Hawking radiation does not occur in this
theory.

\section{\label{discussion}Discussion}

A few points deserve further discussion. \ First of all, although we have
found no Hawking radiation in the limit that $\frac{1}{M \Lambda} \rightarrow
0$, the reader may wonder if it is possible that a normal spectrum will appear
at a higher order in $\frac{1}{M \Lambda}$. \ But in fact this is not
possible. \ For example, if the spectrum actually had a Boltzmann distribution
going like $e^{- \omega / T}$ with $T \propto 1 / M$, then we could keep
$\omega M$ fixed as we take $\frac{1}{M \Lambda}$ to zero, and the Hawking
effect would not disappear. \ Our calculation is perfectly valid in this limit
however, since even with $\omega = 0$ we obtain fluctuations around the
background which satisfy the WKB approximation. \ Note that we can similarly
rule out a spectrum such as $e^{- k / T}$. \ The existence of the non-trivial
$\omega = 0$ solutions is interesting in and of itself- they are indeed time
independent, and one might wonder if they can be thought of as hairs for the
ghost condensate black holes. \ This is actually not the case, since they have
divergent energy momentum at the horizon as seen by freely falling observers.

 We have thus reached the conclusion that the ghost condensate
phonons do not have a thermal spectrum of Hawking radiation, and in
particular, the radiation vanishes in the limit $\frac{1}{2 M} \ll \Lambda \ll
M_{\tmop{pl}}$. \ To this author's knowledge, this is the first known example
of a completely non-thermal Hawking spectrum.

 It seems likely that the ghost condensate is incompatible with
quantum gravity, and it is interesting that signs of this fact show up in the
Hawking spectrum, without requiring any particular couplings to other fields.

 There are a number of outstanding issues which would be
worthwhile to address in future investigations. \ First of all, it would be
nice to show explicitly that the GSL is violated in the ghost condensate
through direct application of its null energy condition violating properties.
\ In particular, one could try to send lumps of net negative energy into a
black hole, shrinking the horizon area while reducing the entropy of the
exterior. \ It would also certainly be very interesting to find theories other
than the ghost condensate in which the Hawking spectrum is abnormal. \ Along
the same lines, one might wonder whether it is possible to violate the GSL
with a fluid which is a small modification of the ghost condensate, perhaps
even one which does not quite violate the null energy condition. \ Work is
ongoing on these various issues.

\section{Acknowledgements}

We would like to thank Ben Freivogel for collaboration at the initial stages
of this project. \ In addition, Sergei Dubovsky, Ted Jacobson and Ami Katz all
provided very useful and interesting discussions. \ This work was supported by
the Department of Energy grant number DE-FG02-01ER-40676.

\appendix

\section{\label{A}Christoffel Symbols}

Here we list the Christoffel symbols for Schwarzschild space-time in the
$\{\tau, r, \theta, \phi\}$ coordinate system defined in the text. \ We have
defined $v (r) \equiv - \sqrt{2 M / r}$ and $v' \equiv d v / d r$. \ All
symbols not listed are zero unless required by symmetry. \
\begin{eqnarray}
  \Gamma^{\tau}_{\tau \tau} & = & v^2 v' \nonumber\\
  \Gamma^{\tau}_{\tau r} & = & - v v' \nonumber\\
  \Gamma^{\tau}_{r r} & = & v' \nonumber\\
  \Gamma^{\tau}_{\theta \theta} & = & v r \nonumber\\
  \Gamma^{\tau}_{\phi \phi} & = & v r \sin^2 \theta \nonumber\\
  \Gamma^r_{\tau \tau} & = & - (1 - v^{2}) v v' \nonumber\\
  \Gamma^r_{\tau r} & = & - v^2 v' \\
  \Gamma^r_{r r} & = & v v' \nonumber\\
  \Gamma^r_{\theta \theta} & = & - (1 - v^2) r \nonumber\\
  \Gamma^r_{\phi \phi} & = & - (1 - v^2) r \sin^2 \theta \nonumber\\
  \Gamma^{\theta}_{\theta r} & = & 1/r \nonumber\\
  \Gamma^{\phi}_{\phi r} & = & 1/r \nonumber\\
  \Gamma^{\phi}_{\phi \theta} & = & 1/\tan \theta \nonumber\\
  \Gamma^{\theta}_{\phi \phi} & = & - \sin \theta \cos \theta \nonumber
\end{eqnarray}

\end{document}